\documentclass[usenatbib,usegraphicx]{aastex}

\usepackage{emulateapj5}

\def\s8{$\sigma_8$\,}
\def\Om{$\Omega_M$\ }
\def\Ome{$\Omega_M$}

\def\mlim{$M_{lim}$\ }
\def\mlime{$M_{lim}$}
\def\tCDM{$\tau$CDM}
\def\lCDM{$\Lambda$CDM}

\def\xray{\hbox{X--ray} }
\def\kev{{\rm\ keV}}

\def\myputfigure#1#2#3#4#5%
{\vskip#5pt\makebox[0pt]{\hskip#2in
\includegraphics[width=#3\textwidth]{#1}}\vskip#4pt\hfill}

\begin{document}

\title{Cosmological Implications of the BIMA 30~GHz \\
Sunyaev-Zel'dovich Effect Galaxy Cluster Survey}

\submitted{Submitted to ApJ June 28, 2002}

\author{Yen-Ting Lin\altaffilmark{1} and Joseph J. Mohr\altaffilmark{1,2}}

\altaffiltext{1}{Department of Astronomy, University of Illinois; 
Urbana, IL 61801; ylin2@astro.uiuc.edu}
\altaffiltext{2}{Department of Physics, University of Illinois; 
Urbana, IL 61801; jmohr@uiuc.edu}

\begin{abstract}

We explore the cosmological implications of 7 deep
survey fields observed at the Berkeley-Illinois-Maryland
Association (BIMA) Array with 30 GHz receivers.
These observations probe the Cosmic Microwave Background 
anisotropy on scales corresponding to $l\sim 5500$, and an earlier analysis of these data detected no galaxy clusters.
We use numerical cluster simulations and mock observations to
characterize the cluster detection efficiency for each of the BIMA
fields.  With these detection efficiencies we derive constraints on
the cosmological parmaters $\Omega_M$ and $\sigma_8$, ruling out
those models which overproduce galaxy clusters.  Using only these
seven BIMA fields, we calculate a 2$\sigma$ upper
limit of $\sigma_8 < 1.00\, \Omega_M^{-0.43 \Omega_M-0.22}$ for flat models
with $0.1\le\Omega_{M}\le1$.  When the power spectrum of density
fluctuations is COBE normalized, 
we find $\Omega_M < 0.63$ at $95\%$ confidence level for flat models.
This constraint includes our estimate of the large uncertainties in the 
SZE flux--virial mass relationship as well as published uncertainties in the Hubble parameter, the COBE power spectrum normalization and the primordial power spectrum index.  In addition, we account for the effects of sample variance.  Thus, we conclude that the non--detections are to be expected in a low $\Omega_M$ universe, given the sensitivity and solid angle of the deepest SZE survey to date.

\end{abstract}
\keywords{cosmology: observation -- galaxies: clusters -- cosmic microwave background}

\section{Introduction}

Cosmic microwave background  (CMB) images provide information about 
the presence of galaxy clusters over a wide range of redshift, because galaxy
clusters leave signatures in the CMB through
the so--called Sunyaev-Zel'dovich effect \citep[SZE;][]{sunyaev70,sunyaev72}.
The SZE is produced by inverse Compton interactions between the
CMB photons and the hot electrons ($k_B T_e \sim 1$\,keV, 
see \S\ref{sec:sze}) in the intracluster medium. The intracluster medium
is shocked and heated in the process of cluster formation.
The net result in the SZE is a transfer of energy from the electron population 
to the CMB photons, leading to a distortion in the CMB spectrum. The
thermal SZE signature of a cluster in 1~cm ($\sim$ 30~GHz) observations is a
reduction in the CMB brightness.  On angular scales of $\sim 5^{'}$ and smaller
the thermal SZE contribution to the CMB anisotropy can dominate that of
the primary CMB anisotropy \citep{holder99,holzapfel00a,hu01,springel01}.

Here we discuss the cosmological implications of 7 deep survey fields observed
at the Berkeley--Illinois--Maryland Association (BIMA) array with the 
30~GHz receivers \citep{carlstrom96,holzapfel00a}.
The observations were made during 1997 and 1998, in a compact configuration 
at 28.5\,GHz, providing a Gaussian primary beam with FWHM$\sim 6'_.6$.
This survey, originally planned to probe the CMB anisotropy on
arcminute scales $l\sim 5500$ (corresponding to angular size $\sim 2'$),
provides relatively large sky coverage ($\sim 250$ arcminute$^2$) for an
arcminute-scale anisotropy experiment.  These observations have been used
to place an upper limit on small scale CMB anisotropy \citep{holzapfel00a}.
Continued observations on these and additional fields has led to a detection
of anisotropy with flat band power at the level $Q_{flat}=6.1^{+2.8}_{-4.8}\mu
K$, where uncertainties describe 68\% confidence regions 
\citep[][see also \citealt{dawson02}]{dawson01}.  
In addition, no galaxy cluster detections were reported, which is interesting because of the sensitivity of SZE surveys to high redshift clusters.  In the currently favored cosmology, these seven fields sweep out as much comoving volume to $z\sim2$ as a 130~deg$^2$ local survey sweeps out to $z\sim0.1$.

In principle, cluster surveys constrain cosmological parameters
through the effects those parameters have on the volume surveyed per
solid angle and the evolution of cluster abundance 
\citep[e.g.][]{haiman01,holder01b,mohr01}.  In this analysis we focus on the lack 
of clusters in these deep '97 and '98 BIMA fields.  A less sensitive, earlier 
survey with the Australian Telescope Compact Array also provided upper limits 
on the CMB anisotropy on similarly small scales \citep{subrahmanyan00}. An 
analysis of the cosmological implications of that survey, which relied on 
the assumption that the dominant source of anisotropy on arcminute scales is 
the thermal SZE from clusters, concluded that a low-$\Omega_M$ universe 
is preferred \citep{majumdar00}.

An analysis of this BIMA survey allows us to rule out cosmological models 
which produce too many clusters.  The expected number of detected clusters in 
a survey is
\begin{equation}
\langle N \rangle  =
        \int d\Omega \int dV
        \int_0^{\infty} dM\,f(M)\,\frac{dn}{dM}\left(M,z\right),
\label{eq:expectation}
\end{equation}
where $(dn/dM)dM$ is the comoving number density of
objects of mass between $M$ and $M+dM$, $f(M)$ is the detection efficiency
($0\le f(M)\le 1$), which takes into account the fact that only clusters with
masses above some mass limits can be detected.
The first integral is over the solid angle of the survey, the second integral
is over the redshift, and the third integral sums over the portion
of the cluster mass function to which our survey is sensitive.
In addition to its dependence on mass, the detection efficiency $f(M)$ depends 
on the sky position $(\alpha,\delta)$, the redshift $z$ and
cosmological parameters such as $H_0$, $\Omega_M$ and $\Omega_\Lambda$.

Cosmological constraints come through comparison of the observed and 
the expected number of clusters.  In the BIMA survey no clusters were 
detected; assuming Poisson statistics, the probability $P$ of observing $0$ 
clusters is $ P(0|\langle N\rangle) = e^{-\langle N\rangle}$.  Using this 
approach we explore the parameter space spanned by $\Omega_M$ and $\sigma_8$, 
checking the consistency of these cosmological models.  We also consider the
uncertainties in the cluster detection efficiency $f(M)$, Hubble parameter,
COBE power spectrum normalization, power spectrum index and the effects of 
sample variance.

This paper is organized as follows: in \S\ref{sec:masssens} we present a 
general description of interferometric SZE observations, describe the BIMA 
survey data, and determine the survey detection efficiency via mock 
observations of hydrodynamic cluster simulations. We discuss the ingredients 
needed for the calculations of cluster abundance in \S\ref{sec:yields}. 
The cosmological constraints and a discussion on effects of systematic errors
appear in \S\ref{sec:res}. We present a summary and discussion on prospects of
future interferometry SZE observations in \S\ref{sec:summary}.
We discuss the $H_0$ scaling of interferometric SZE observations in the
Appendix.

\section{Survey Sensitivity to Clusters}
\label{sec:masssens}

The expected number of detected clusters 
in each cosmology depends sensitively on the detection efficiency function
$f(M)$, characterized by a limiting detectable mass $M_{lim}$ at which
$f(M_{lim}) = 0.5$.
We determine the detection efficiency by conducting mock observations of
hydrodynamical cluster simulations using the exact characteristics of
the real BIMA observations.
In this section, we first discuss the interferometric observations of
the cluster SZE, then describe the BIMA data, and provide a detailed 
description of the determination of the detection efficiency.

\subsection{Interferometric SZE Observations of Clusters}
\label{sec:sze}

Interferometric observations of cluster SZE have greatly advanced in the
past decade \citep{carlstrom00}.
Interferometers can observe a wide range of angular scales
simultaneously, reduce possible confusion due to
radio point sources
toward galaxy clusters, and generate high-fidelity SZE images.

Interferometers measure the visibility $V_{\lambda}(u,v)$, 
whose Fourier transform 
pair, the specific intensity $I_{\lambda}(\theta_{x}, \theta_{y})$,
is directly proportional to the Compton $y(\theta_{x},\theta_{y})$ distribution
of the cluster. Along any line of sight,
\begin{equation}
y=\sigma_T\,\int\,dl\, n_{e}{k_{B}T_{e}\over m_{e}c^{2}},
\label{eq:compton-y}
\end{equation}
where $\sigma_T$ is the Thomson scattering cross-section, 
$n_{e}$ is the electron number density, $m_{e}$ is the electron 
rest mass, $c$ is the speed of light, $k_B$ is the Boltzmann constant 
and $T_e$ is the electron temperature.  
The SZE temperature 
decrement is then $\Delta T=g(x)\,y\,T_{CMB}$,
where $x=h\nu/k_{B}T_{CMB}$ is the dimensionless frequency, $T_{CMB}=2.728$~K 
\citep{fixsen96} is the mean cosmic microwave background temperature,
and, within the Kompaneets approximation, 
$g(x) = x(e^{x}+1)/(e^{x}-1)-4$ is the amplitude of the spectral distortion.
In the Rayleigh--Jeans regime ($x<<1$) $g(x) \approx -2$, and at
the central frequency of these BIMA
observations, $\nu=28.5$~GHz, $g(x)=-1.9583$.  
We convert this decrement distribution $\Delta T(\theta_{x},\theta_{y})$ 
into specific intensity $I_\lambda(\theta_{x},\theta_{y})$ using
\begin{equation}
I_\lambda (\theta_{x},\theta_{y}) = 
	2\,{k_{B}\Delta T(\theta_{x},\theta_{y})\over\lambda^2}.
\label{eq:intensity}
\end{equation} 
where $\lambda$ is the wavelength of observation.
The visibility $V_\lambda(u,v)$ then is the Fourier transform of the intensity
distribution or image 
$I_{\lambda}(\theta_{x},\theta_{y})$ multiplied by the telescope primary beam
$B(\theta_{x},\theta_{y})$,
\begin{eqnarray}
\nonumber
V_\lambda (u,v) & = & \int\!\!\!\int d\theta_x d\theta_y\,\, I_{\lambda}\,
	B\, e^{-2\pi i(u\theta_x +v\theta_y)}	\\
       & = & \tilde{I_{\lambda}} \otimes \tilde{B},
\label{eq:visibility}
\end{eqnarray}
where $\tilde{I_{\lambda}}$ and $\tilde{B}$ are the Fourier transforms of 
$I_\lambda$ and $B$, respectively, and $\otimes$ denotes convolution. 
For this analysis, we take the BIMA primary beam to be a Gaussian with
FWHM\,$=6'.33$; therefore $\tilde{B}$ is also a Gaussian. 
Because $V_\lambda(0)=\int d^2\theta\,I_\lambda B$, 
the overall normalization of the measured visibilities in
the case that the cluster lies well 
within the BIMA primary beam (i.e. $B\simeq1$) and observation frequencies
are in the Rayleigh-Jeans limit is directly related to the cluster flux
within the virial region:
\begin{equation}
\nonumber
V_\lambda (0) \ \stackrel{B \simeq 1}{\longrightarrow} \ S_{200}  = 
        \frac{-4\,\sigma_T\,k_B^2\,T_{CMB}}{\lambda^2\,m_e c^2\,d_A(z)^2}
	\langle T_e \rangle\,
        \frac{f_{ICM} M_{200}}{\mu_e\,m_p}
\label{eq:flux}
\end{equation}
where $\langle T_e \rangle$ is the electron density weighted mean temperature 
of the cluster, $d_A$ is the angular diameter distance,
$\mu_e$ is the mean molecular weight per electron, $m_p$ is the rest
mass of proton, $f_{ICM}$ is the ratio of intracluster gas mass to binding
mass, and $M_{200} \equiv 4\pi/3\, r_{200}^3 \times 200 \rho_c$ is a measure of
total cluster binding mass, within which the mean overdensity is 200 times of 
the critical density $\rho_c$. This expression also defines the virial radius 
$r_{200}$.  Note that Eqn~\ref{eq:flux} does not account for any
SZE flux contribution from outside the cluster virial region.
Also notice that, the only cosmology--dependence of the flux comes from $d_A$.

To examine the sensitivity of cluster visibilities to cluster size and
morphology, we adopt a cluster
temperature decrement model. We use the spherical $\beta$ model
\citep{cavaliere78}, 
$\Delta T(\theta) = \Delta T_0\,(1+(\theta/\theta_{c})^2)^{(1-3\beta)/2}$,
where $\Delta T_0$ is the temperature decrement toward the cluster center,
and $\theta_{c} = r_c/d_A$ is the angular core radius.  Comparison of this 
model with our simulated cluster $\Delta T$ maps 
indicates that it provides a reasonable
description of clusters with $\left<\beta\right>\simeq1.1$.  
For $\beta = 4/3$ in the case where the primary beam is larger
than the cluster (i.e. $B\simeq1$), the visibility and the SZE flux within the
virial radius is related by:
\begin{equation}
V_\lambda(q)  = 
	{S_{200}\over \Theta} e^{-2\pi \theta_c q} 
\label{eq:VandS}
\end{equation}
where $q = (u^2+v^2)^{1/2}$,
$\Theta=1-(1+\left({\theta_{200}/\theta_c}\right)^2)^{-1/2}$ 
is the fraction of the total cluster flux that lies within 
$\theta_{200} = r_{200}/d_A$, the angular extent of the virial radius.

In general, all interferometers are incapable of measuring $V_\lambda(0)$ 
directly,  and the BIMA interferometer is constrained to measure cluster 
visibilities at $q>D/\lambda\sim600$ ($D$ is the telescope diameter). 
Eqn~\ref{eq:VandS} shows clear that the shape, the size and the
total integrated flux from the cluster affect the measured visibilities.
An accurate determination of the survey mass selection function $f(M)$ 
requires knowledge of cluster morphologies.  This fact also means that
the simple Hubble parameter $h$ scaling that one derives by assuming 
that the total cluster SZE flux is the only important variable is 
incorrect.  We discuss this point more in the Appendix. In
future surveys it will be possible to use observed cluster visibilities
along with independent mass measurements to accurately determine the
survey selection function.  In the current work we use hydrodynamical
simulations of clusters to determine the survey selection.

\begin{table*}[htb]
\begin{center}
\caption{Basic description of the fields \tablenotemark{\dagger}}
\begin{tabular}{lccccccc}
\hline \hline
  &\multicolumn{2}{c}{Pointing Center (J2000)} & Beam Size & Time &rms &rms \\
Field \tablenotemark{\star} & $\alpha$ & $\delta$ & ($'' \times ''$) & (hr) & (
$\mu$K) & ($\mu$Jy bm$^{-1}$) \\
\hline
p1643 & 16 45 11.3        & +46 24 56       & 98.3$\times$116.1 & 43.1 & 25.1 &
 191 \\
VLA1312 & 13 12 17.4      & +42 38 05       & 95.2$\times$113.4 & 35.5 & 31.0 &
 223 \\
pss0030 & 00 30 16.4      & +17 02 40       & 99.1$\times$115.5 & 36.6 & 26.6 &
 203 \\
BF0028 & 00 28 04.4       & +28 23 06       & 108.9$\times$118.8 & 77.6 & 11.5
& 99 \\
HDF\_BIMA & 12 36 49.4    & +62 12 58       & 110.9$\times$122.0 & 32.0 & 18.6
& 168 \\
BF1821 & 18 21 00.0       & +59 15 00       & 108.4$\times$122.4 & 43.5 & 22.4
& 224 \\
BF0658 & 06 58 45.0       & +55 17 00       & 105.2$\times$148.0 & 44.7 & 29.3
& 304 \\
\hline
\multispan{4}{$^\dagger$From \citet{holzapfel00a}}\hfil\\
\multispan{4}{$^\star$Field names are refered in \citet{holzapfel00a} as
BDF1--7, respectively}\hfil\\
\end{tabular}
\end{center}
\vskip-20pt
\end{table*}

\subsection{The BIMA Deep Survey Data
\label{sec:data}}

The data are from a study of the CMB anisotropy on
arcminute angular scales \citep{holzapfel00a}.  Fields were chosen to contain
no known, bright radio point sources.  The observations were made 
with the BIMA array with a central frequency of $28.5$ GHz and $\sim$800~MHz
of bandwidth. The receivers are based on low-noise, high electron
mobility transistor (HEMT) amplifiers, and are sensitive only to 
right-circularly polarized radiation. More details on the receivers are given
in \citet{carlstrom96}. Supporting observations were done with the 
Owens Valley Radio Observatory (OVRO) for aid in the detection and removal of 
radio point sources in two of the fields.

Of the seven BIMA fields included in this analysis, three were observed
in 1997 (the first 3 in Table 1), 
and the rest were observed in 1998. Additional observations of these and other
CMB fields have since been obtained \citep{dawson01,dawson02}.
Previous observations by other teams \citep{jones97,richards97} suggested 
temperature decrements with marginal significance in two of the
1997 fields.  However, the higher sensitivity BIMA observations failed
to confirm these detections.  The four 1998 fields are evenly distributed in 
right ascension, in regions of low IR dust emission that are free of bright 
optical and X-ray sources. Additional details of these fields are summarized 
in Table 1 and in \citet{holzapfel00a}.

For the purposes of our analysis, only the list of visibility measurement 
positions $(u_i,v_i)$ and uncertainties $\sigma^2_i$ are required.  We perform 
no further analysis of the actual observations;  however, we describe
some general details of the analysis from \citet{holzapfel00a}.
To remove point sources from the data, a long baseline ($> 2.4\,k\lambda$) map 
was made for each field; five point sources were found in the seven fields 
(but see \citealt{dawson02}). 
Identified point sources were removed from the visibility data by subtracting 
the best fit model of the point source, modulated by the primary beam.

The processed data were analyzed by a Bayesian maximum likelihood method.
A likelihood function was constructed from the visibility correlation matrix.
With this technique the consistency of a theory on CMB fluctuations could be 
obtained by comparison to the data.  Because the fields are independent, the 
total probability
is the product of that from individual fields. The joint likelihood for the
combined data is expressed in terms of the flat-band power amplitude:
$Q_{flat} \equiv 5\,C_l\,l(l+1)/24 \pi < 14.1 \mu$K at $95\%$ confidence,
where $C_l$ is the amplitude of the CMB angular power spectrum of mode $l$.

\citet{holzapfel00a} found that, of the seven fields, two show excess power 
($Q_{flat}>0$); however, the confidence of a  non-trivial signal is only $44\%$ for the joint likelihood analysis, and each  of the fields is consistent with no signal at $95\%$ confidence level.  Based on this information together with the argument that the thermal SZE is the dominant source of anisotropy on this angular scale, we infer that no galaxy clusters were 
detected at or above the 95\% confidence limit.  This inference is uncertain, because quantitative results of a sophisticated search for extended, negative sources (i.e. cluster SZE signatures) have not been reported by the authors in the three manuscripts that have appeared to date.  Thus, we also examine the sensitivity of our limits on $\Omega_M$ to the numerical significance we ascribe to the non--detections (see $\S$\ref{sec:select}).  

\subsection{Determining the Detection Efficiency}
\label{sec:mock}

The cluster mass $M$ is the single most important 
factor in interferometric cluster SZE observations (e.g. \citealt{holder00}).
From Eqn~\ref{eq:flux} we expect a direct correspondance
between the flux from a cluster and its mass. In an ideal
flux limited survey, 
this naturally leads to the functional form of the
detection efficiency which is a step function: $f(M \ge M_{lim}) =1$ and
$f(M < M_{lim}) =0$, where $M_{lim}$ is the cluster mass that
corresponds to the minimum detectable flux of the survey.  However,
we expect scatter about the total SZE flux-- virial mass relation, because
of departures from equilibrium and variations in cluster morphology at a
fixed mass.  These morphological variations are particularly important in an
interferometric survey, where the observed visibility has direct sensitivity
to cluster shape and size.  Therefore, we use an $f(M)$, that varies continuously
from $f(M)=0$ for $M<<M_{lim}$ to $f(M)=1$ for $M>>M_{lim}$.

The method is that given a set of cosmological parameters 
(here $H_0$, $\Omega_M$ and $\Omega_\Lambda$), a redshift, and a position on 
the sky $(\alpha,\delta)$, we simulate observations and detections of an 
ensemble of clusters with a large range of masses.  These mock interferometric 
SZE observations have characteristics such as sensitivity, $u$-$v$ coverage,
and primary beam identical to those carried out in the BIMA survey.
We then choose the detection efficiency $M_{lim}$ so that it
reflects the mass of clusters detected at 95\% confidence.

In practice, we account for the cosmological sensitivity of the limiting mass
through the dependence of the cluster flux on the angular diameter distance 
$d_A$ (Eqn~\ref{eq:flux}).  To determine the coordinate dependence of \mlim 
for each BIMA field, we carry out sets of mock observations at 8 off-axis
angles $\theta$ in the range $0'\le\theta\le3'.5$ and 6 azimuthal angles 
$\phi$ (although we expect azimuthal symmetry). At each of these field
positions $(\theta,\phi)$, we examine the detectability of 48 simulated 
clusters (observed along 3 orthogonal axes, \S\ref{sec:hydro}) 
at redshift $z=0.5$, where the cluster angular diameter distance is 
determined within a fiducial cosmology.
To determine the redshift variation of the limiting mass
within each BIMA field, we make mock, on--axis ($\theta=0$) observations of these 
48 clusters at different redshifts.  The mass limit at redshift $z$ is determined using simulated clusters output at that same redshift.

For each of these mock observations, we convert the Compton $y$ map determined
from the hydrodynamic simulations to visibilities using a Fast 
Fourier Transform (FFT) according to the prescription 
outlined in \S\ref{sec:sze}.  Visibilities are then sampled from this map
at the same locations in $u$-$v$ space probed by the BIMA observations in each 
field.  We add Gaussian noise to the visibility that is appropriate given the 
uncertainty of that same visibility in the BIMA observations; in fact, we
scale these uncertainties down by a factor, which corresponds to a longer
observation with BIMA (using $\sigma_V\propto1/\sqrt{t_{obs}}$).  This 
reduction in noise allows a more accurate estimate of the cluster mass
corresponding to a particular detection significance 
(see Fig~\ref{fig:deltachi}), and this scaling can be
removed after the detection.  We have tested the accuracy with which one
can introduce these longer observing times and then remove them
after the mock detection by calculating the limiting mass for a wide
range of observing time scale factors.  Results indicate that the
limiting mass corresponding to a 2$\sigma$ detection 
is independent of the exact scale factor used (assuming it is
large) at the 1\% level.

\subsubsection{The Hydrodynamical Simulations}
\label{sec:hydro}

The cluster models used in the mock observations are from an ensemble of
48 hydrodynamical cluster simulations in four Cold Dark Matter (CDM) 
cosmologies:
(1) Standard CDM (SCDM: $\Omega_{M}=1$, $\sigma_8=0.6$, $h=0.5$,
$\Gamma=0.5$), (2) Tilted CDM (\tCDM: $\Omega_{M}=1$, $\sigma_8=0.6$, $h=0.5$,
$\Gamma=0.24$), (3) Open CDM (OCDM: $\Omega_{M}=0.3$,
$\sigma_8=1$, $h=0.8$, $\Gamma=0.24$),
and (4) Lambda CDM (\lCDM: $\Omega_{M}=0.3$, $\Omega_{\Lambda}=0.7$,
$\sigma_8=1$, $h=0.8$, $\Gamma=0.24$).
Here $h$ is the Hubble parameter in units of 100 km/s/Mpc.
The initial conditions of the simulations are Gaussian random fields
consistent with a CDM transfer function with index $\Gamma$ \citep{davis85}.

The simulation scheme is particle-particle-particle-mesh smoothed-particle
hydrodynamics \citep[P3MSPH]{evrard88},
with fixed intracluster gas mass fraction
$f_{ICM}=0.2$ (but we correct to $f_{ICM}=0.12$, consistent
with observations for $h=0.70$; \citealt{mohr99}).
Radiative cooling and heat conducting are ignored.
The 48 simulated clusters have similar fractional mass resolution; the spatial
resolution varies from 125--250~kpc.
The masses of the final cluster sample vary by an order of magnitude.
More detail on the simulations can be found in other analyses that
have employed these same simulations 
\citep{mohr97a,mohr99,mathiesen99,mathiesen01}.

Following procedures described in \citet{evrard90}, we create
Compton $y$ parameter images along three orthogonal lines of sight for each
cluster.  For these calculations each cluster is imaged as though it
were at a redshift $z=0.06$.  These projections serve as templates for the mock
observations used to determine the limiting masses.  In determining the
limiting mass at a redshift $z$, we use the simulation outputs that
correspond to that redshift; this ensures that our mass limit includes
the effects of cluster structural evolution with redshift.

\myputfigure{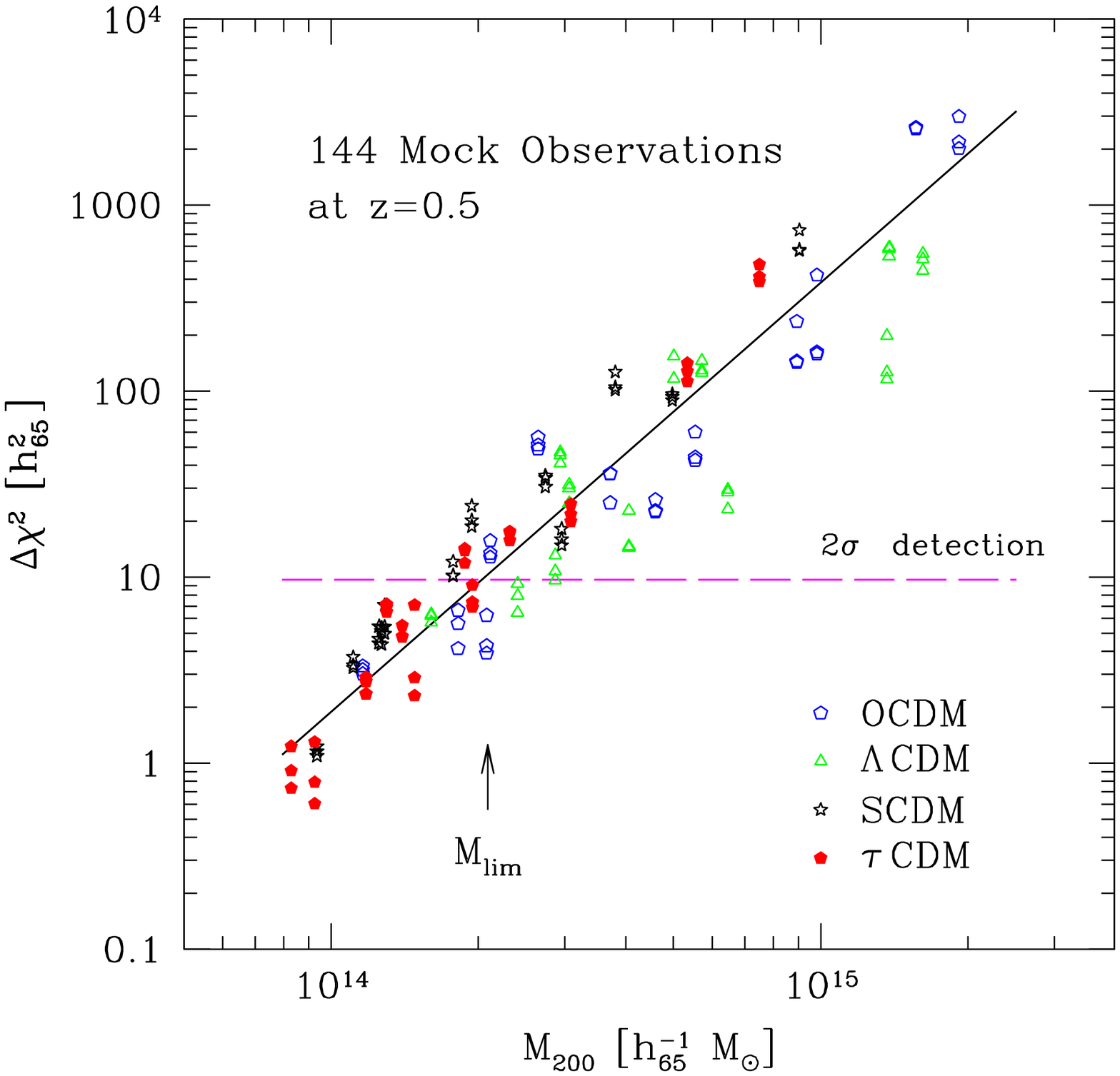}{3.2}{0.5}{-30}{-0}
\figcaption{The detection significance, measured as $\Delta \chi^2$,
of mock observations of the 48 simulated clusters in one BIMA field (BF0028).
Each cluster is observed along three orthogonal lines of sight. The 
intersection of the best-fit line with the  $\Delta \chi^2 = 9.7$ line
(corresponding to $2\sigma$ detection) determines the detection threshold mass 
\mlime. Here and afterwards $h_{65} \equiv h/0.65$.
\label{fig:deltachi}}
\bigskip

\subsubsection{Mock Cluster Detection}
\label{sec:detect}

We analyze the mock observations using software developed to 
analyze real cluster observations \citep{reese00}.  For 
simplicity and fast computation, we model the ICM 
density distribution using a spherical $\beta$ model \citep{cavaliere78}
with $\beta = 4/3$. For this value of $\beta$,  the Fourier conjugate of the
model is an exponential (see Eqn~\ref{eq:VandS}); additionally, this value is roughly consistent with the best fits to the SZE decrement images from the simulations ($<\beta >=1.1$).  Furthermore, we do not include
the effects of the primary beam, and this allows us to bypass the FFT
normally required in each fit iteration.  This fast detection scheme should
be appropriate near the detection threshold, where there is little information
about the detailed shape of the cluster.  

For each cluster we fit the core radius and the central
decrement simultaneously by minimizing the $\chi^2$ difference between
the fit and the observations.  We do not actually simulate ``finding''
the clusters within the field;  rather, we assume that the approximate
location of each cluster can be obtained by examining the BIMA observation
in the image plane.  We are currently exploring other techniques, which
would allow detection and characterization of clusters in the Fourier domain.
The $\chi^2$ difference between the null model and the best fit model
($\Delta \chi^2 \equiv \chi^2_{null}-\chi^2_{fit})$
serves as a measure of the detection significance.
At a point on the field $(\theta,\phi)$ and a redshift $z$, we repeat this 
procedure for all 48 cluster models (viewed along three axes) and obtain a 
tight correlation between cluster mass $M_{200}$ and detection significance 
$\Delta \chi^2$.
We determine the best-fit of the $\Delta \chi^2-M_{200}$ relation by
a downhill simplex method \citep{press92}.  For 2 degrees of freedom the 95\% 
confidence limit corresponds to $\Delta \chi^2 = 9.7$. 
The limiting mass $M_{lim}(\theta,\phi,z)$ is the intersection of the
best-fit line with the $\Delta \chi^2 = 9.7$ line, and the detection
efficiency function $f(M)$ is constructed from the scatter about the best-fit 
line.  Fig~\ref{fig:deltachi} is a plot of $\Delta\chi^2$
versus $M_{200}$ for mock observations on field BF0028, observed on--axis
$(\theta = \phi =0)$ at $z=0.5$.

\myputfigure{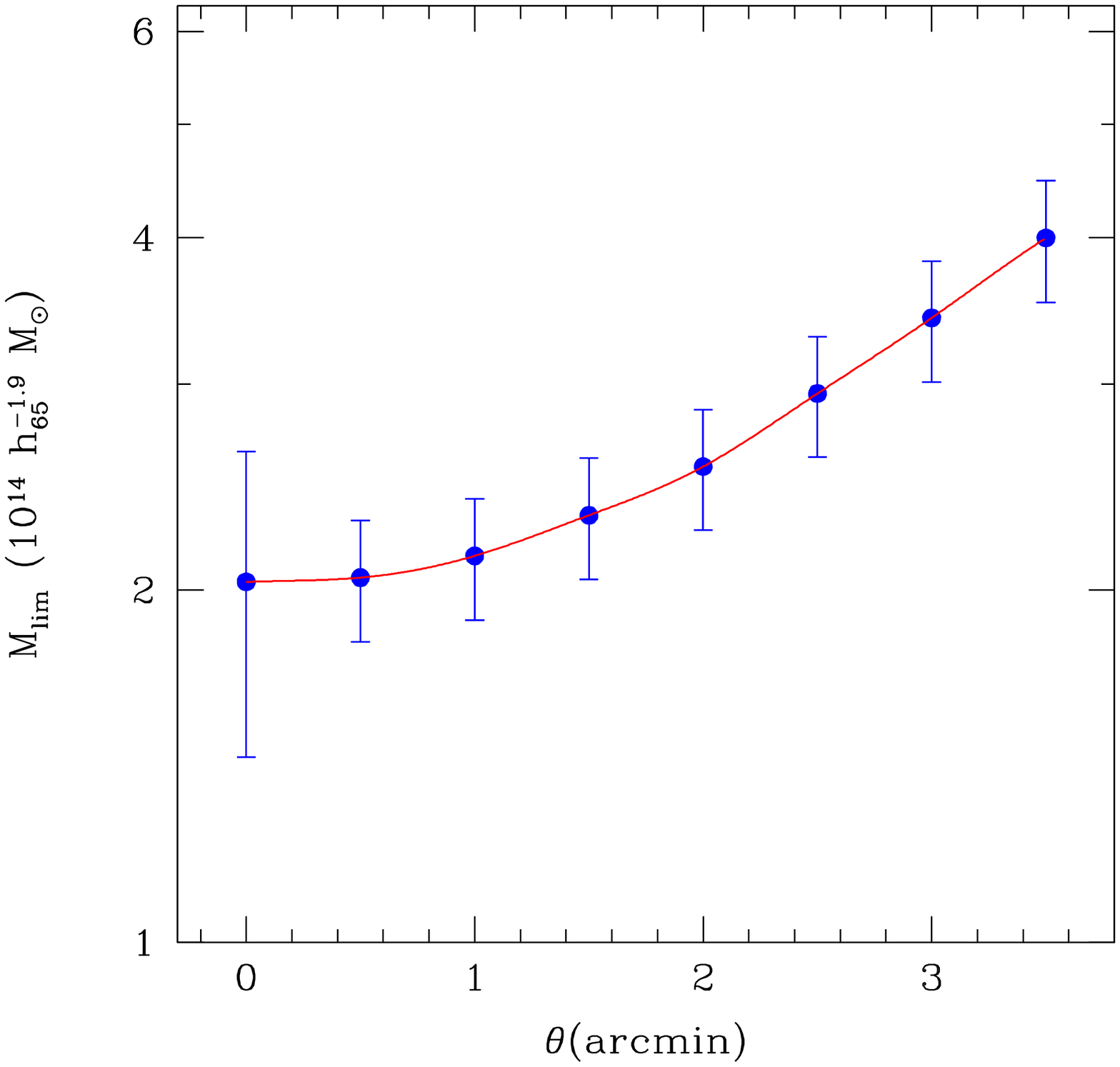}{3.2}{0.5}{-20}{-0}
\figcaption{The angular dependence of the limiting mass in the field
BF0028.  These limiting masses are determined using mock
observations of simulated galaxy clusters.  The mock observations
mimic the actual observations of each BIMA field.  Also shown is the spline,
which we use to determine \mlim\ for arbitrary off-axis angle.
(For the $h$ dependence of \mlim see the Appendix.)
\label{fig:theta}}\vskip5pt

We have compared the limiting mass given by the faster detection 
algorithm described above (fixed $\beta=4/3$ model and no primary beam) 
with that obtained from full $\beta$--model fitting that includes the 
effects of the primary beam.  The differences in the derived 
limiting mass \mlim are at the 1\% level.

The {\it{rms}} scatter with respect to the $\Delta \chi^2-M$ correlation 
gives the uncertainty in the limiting mass and determines $f(M)$.
We model the detection efficiency by assuming Gaussian scatter in the
$\log{\Delta \chi^2}-\log{M}$ relation.  
Taking $\lambda=\log{\Delta\chi^2}$ and the {\it rms} scatter
in $\lambda$ to be $\sigma_{\lambda}$, we can write the
detection efficiency as
\begin{equation}
f(M) = {1\over\sqrt{2\pi\sigma^2_{\lambda}}}
	\int_{\lambda_{min}}^\infty \!\!\!\!\!\!\!\!\!
	d\lambda \, e^{-{1\over2}\left(
	{\lambda-\overline{\lambda(M)}\over\sigma_\lambda}
	\right)^2}
\label{eq:fM}
\end{equation}
where $\lambda_{min}=\log{9.7}$ (i.e. a 2$\sigma$ detection) and 
$\overline{\lambda(M)}$ is the best fit $\log\Delta\chi^2$ for mass $M$.

Applying the same procedure for each deep field,
we obtain $M_{lim}(\theta,\phi,z)$ and $f(M)$ at specific values
of $(\theta,\phi,z)$ for all fields. Values of $M_{lim}$
for arbitrary $(\theta,\phi,z)$ require an interpolation scheme.
We take the limiting mass to be 
$M_{lim}(\theta,\phi,z)=M_{lim}(\theta,\phi,z=0.5) \times r(z)$,
where $M_{lim}(\theta,\phi,z=0.5)$ encodes the angular dependence of
the limiting mass at redshift $z=0.5$, and 
$r(z)$ is the redshift behavior of the mass limit at $\theta=\phi=0$,
normalized so that $r(z=0.5)=1.0$.

\myputfigure{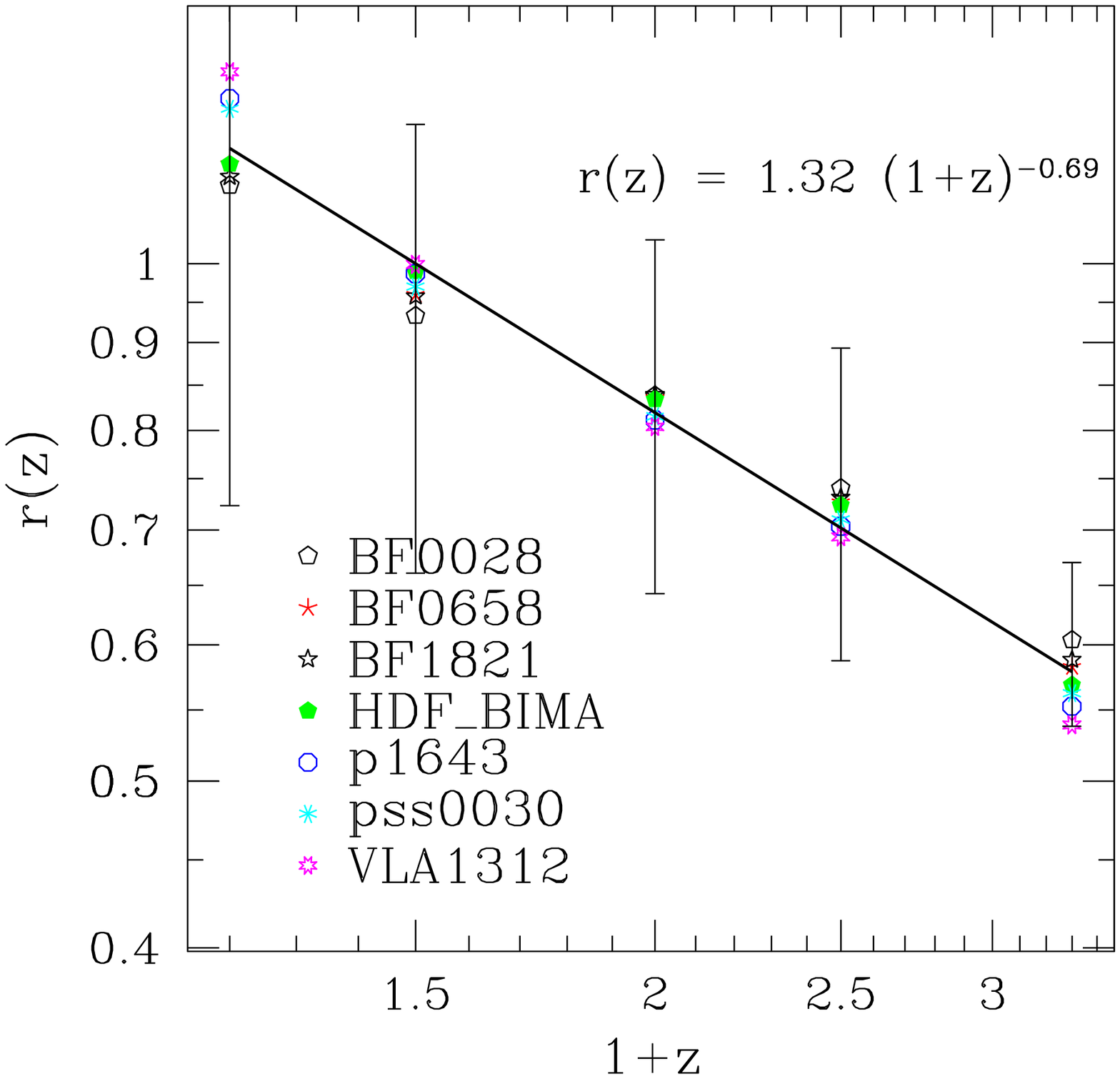}{3.2}{0.5}{-20}{-0}
\figcaption{The redshift dependence of the limiting mass.
Points for each BIMA field denote the measured limiting masses at 5 redshifts
relative to the value at $z=0.5$.
The best fit power law $r(z)$ to the ensemble is noted in the figure.
Error bars are shown only for field BF0028.
\label{fig:redshift}} 
\bigskip

To determine the angular variation of the limiting mass, we average the values 
of $M_{lim}(\theta,\phi,z=0.5)$ over the azimuthal angles $\phi$ that are 
evenly spaced in the $360^{\circ}$ circle; we sample off--axis angles
$\theta$ at $\theta$ = $0'_.5$, $1'_.0$, $1'_.5$, $2'_.0$, $2'_.5$, $3'_.0$ 
and $3'_.5$, where the maximum $\theta$ is close to the HWHM of the beam.
The limiting masses at arbitrary off--axis angles
are calculated use a spline.  Fig~\ref{fig:theta} contains a plot of the 
angular dependence of \mlim in BF0028.

We determine the redshift dependence $r(z)$ using the
limiting mass measurements at redshifts
$z=0.2,\, 0.5,\, 1.0,\, 1.5\ {\rm and}\  2.3$. Fig~\ref{fig:redshift}
contains a plot of the on--axis limiting mass $M_{lim}$ for each field
at these five different redshifts.
A single power--law provides an adequate description of
the redshift behavior of the limiting mass in all seven
fields.  The scatter of the limiting masses around this best fit power--law is
comparable to or smaller than the estimated uncertainties in these limiting masses.

We have also examined the angular variation of the scatter  $\sigma_\lambda$ 
for all off-axis  positions sampled (at $z=0.5$).  The variation is small
(e.g. the {\it rms} variation of $\sigma_\lambda$ in the
field BF0028 is $6.6\%$).
There is also some redshift variation of $\sigma_\lambda$ for
each field (at $\theta=\phi=0$). With a large sample of simulated clusters
it will be possible to better determine the scatter in detectability
$\sigma_\lambda$, but for this analysis we adopt a single
$\sigma_\lambda$ for each field.  We use this in determining
$f(M)$ at all off--axis angles $\theta$ and redshifts $z$.

Finally, as previously noted, the cosmology dependence in \mlim is from the 
angular diameter distance to the clusters. Because we probe a wide range of 
cosmological models, while our limiting mass is derived from a fiducial 
cosmology with $(\Omega_M, \Omega_\Lambda) = (0.3, 0.0)$, we rescale 
\mlim to be consistent with the model considered. Following \citet{haiman01},
the limiting mass in a given cosmology at redshift $z$ is $M_{lim}(z) h = 
M_{lim}^*(z) h^* (\tilde{d_A}(z)/\tilde{d_A}^*(z))^{1.2}$, where
quantities with an asterisk
denote that in the fiducial cosmology, and $\tilde{d_A}(z) \equiv h\,d_A(z)$ 
is the Hubble-parameter-independent part of the angular diameter distance.  
We discuss the Hubble parameter dependence of the survey in the Appendix.

\section{Expected Yields for Any Cosmology}
\label{sec:yields}

From the discussions in \S\ref{sec:masssens} it is clear that in calculating
the expected number of clusters (Eqn~\ref{eq:expectation}), \mlime, $f(M)$ and 
the functional form of the mass function are important ingredients.
For the mass function we adopt the fitting formula given by \citet{jenkins01},
obtained from large-scale N--body simulations of structure formation:
\begin{equation}
\frac{dn}{dM}(M,z) = -0.315\frac{\rho_0}{M}\sigma^{-1}
                 \frac{d \sigma}{d M}\,{\rm{e}}^{-|\,0.61-\ln\sigma\,|^{3.8}}
\label{eq:Jenkins}
\end{equation}
where $\rho_0$ is the present-day matter density and $\sigma(M,z)$ is {\it rms}
variation in overdensity on mass scale $M$ and at redshift $z$.
This fitting formula provides a more accurate abundance of collapsed halos
than does the classical Press--Schechter mass function \citep{press74}. 
In both Jenkins and Press-Schechter mass functions, 
the form and amplitude of $\sigma(M)$ is one essential component. 

We calculate $\sigma(M)$ by directly integrating the power spectrum of density fluctuations,
masked by the Fourier conjugate of a spherical top--hat window function 
$W(k,R)$ over wavenumber $k$, where $R$ denotes the radius of the spherical
top--hat.  We assume the initial power spectrum is of the form
$P_{ini}(k)\propto k^{n}$, which can be related to the power spectrum $P(k,z)$ 
at any redshift after the epoch of matter--radiation equality
by the CDM transfer function $T(k)$ and the growth function $D(z)$.
We use the transfer function given by \citet{eisenstein98}, and
the baryon density parameter $\Omega_B = 0.019\,h^{-2}$ which derives from
deuterium abundance measurements \citep{burles98}. We write 
$P(k, z) \propto k^{n}\,T^2(k)\,D^2(z)/D^2(0)$, 
where $D(z) = g(\Omega(z))/1+z$ and $g(\Omega(z))$ is the 
growth--suppression factor given by \citet{carroll92}. 

We employ a variety of strategies for normalizing $\sigma(M)$.
These include (1) matching the amplitude of the power spectrum
on the largest scales to the constraints from cosmic microwave
background anisotropy measurements \citep[e.g.][]{bunn97,majumdar00},
(2) reproducing the local cluster abundance (e.g., \citealt{viana99}),
and (3) simply adopting the normalization as a free parameter
to be constrained using the BIMA deep field observations.
The first approach probes the power spectrum on much larger scales than
the last two.

Using the \citet{jenkins01} expression for the mass function $dn/dM$ requires
adopting their halo mass definition.  They use
$M_{SO180}$ masses, corresponding to
the mass enclosed within the region that has a mean
overdensity of $180$ times the {\it{background}} density.
The limiting masses we estimate using mock
observations of hydrodynamic simulations are all $M_{200}$'s--- that is, masses
enclosed in regions with mean overdensity of 200 with respect to the
{\it{critical}}
density.  We convert from $M_{200}$ to $M_{SO180}$ using the ``universal''
halo density profile with a concentration parameter $c=5$ \citep{navarro97}.
Any differences between real cluster halos and our adopted model will
manifest themselves as a systematic error in the mass selection function
$f(M)$.  We account for these systematics in our analysis
(see \S\ref{sec:cobe}).

With these ingredients, we calculate the expected number of detected clusters 
$\langle N \rangle$ in our BIMA fields. 
Rewriting Eqn~\ref{eq:expectation} to reflect our approach,
the expected number of clusters detected in a given field is:
\begin{equation}
\langle N \rangle =  
	\!\!
	\int_0^{\theta_{max}}
        \!\!\!\!\!\!\!\!\!\!\!\!
        d^2\theta
	\!\!
        \int_0^{z_{max}}
        \!\!\!\!\!\!\!\!\!\!\!\!
        dz\, d_A^2 {c\,(1+z)^2 \over H(z)}
	\!\!
        \int_0^{\infty}
	\!\!\!\!\!
        dM f \frac{dn}{dM},
\label{eq:comb}
\end{equation}
where 
$\theta_{max}=3'_.5$, $f=f(M,\theta,z)$ is the detection efficiency and
$dn/dM$ is the cluster abundance given in Eqn~\ref{eq:Jenkins}.
We truncate the redshift integration 
at some maximum redshift $z_{max}$, beyond which the cluster formation and
evolution is uncertain. For $\Omega_M = 1.0$ models, $z_{max}$ is set to be
$4$, and for all other models considered $z_{max}=10$. 

\section{Results}
\label{sec:res}

We use the techniques described above to constrain cosmological parameters
with the BIMA deep survey fields.  We restrict our analysis to two sets of
models:  (1) those with purely dark matter ($\Omega_M \le 1$ \& 
$\Omega_\Lambda = 0$, {\it{open}} models hereafter) and (2) those
with dark matter and a cosmological constant ($\Omega_\Lambda\ge0$ and $w=-1$,
where $w \equiv p/\rho$ characterizes the equation of state)
that are geometrically flat ($\Omega_M+\Omega_\Lambda=1$, {\it{flat}} models 
hereafter).

Let us denote any cosmological model to be considered as 
$\{\Omega,\sigma_8\}_j$.  To determine the probability of consistency of this
model with our data, we calculate the expected number of detected
clusters $\langle N \rangle_i$ for each BIMA field $i$.  
Suppose the detection of clusters is a Poisson random process,
the probability of observing zero clusters in field $i$ is then
$P_i(\{\Omega,\sigma_8\}_j) = e^{-\langle N \rangle_i}$.
The total probability 
of observing zero clusters in each of the seven fields is then 
$P_{tot}(\{\Omega,\sigma_8\}_j) = \prod_{i} P_i$.

Below we explore the cosmological information contained in the BIMA fields
themselves ($\S$\ref{sec:s8free}), and then we examine constraints from the 
BIMA fields together with the local, massive cluster abundance 
(\S\ref{sec:local}) and
together with the CMB constraints on the power spectrum normalization
(\S\ref{sec:cobe}). We end this subsection with a discussion of our approach to including the uncertainties in the survey limiting mass, the Hubble parameter, the COBE normalization parameters, and the effects of sample variance.

\subsection{Using Only BIMA Deep Fields}
\label{sec:s8free}

Fig~\ref{fig:bima} contains the cosmological constraints on parameters
\s8 and \Om from the BIMA survey fields alone.  Solid lines show the
equivalent 1$\sigma$ and 3$\sigma$ Gaussian probabilities for flat models,
and dashed lines show similar confidence regions for the open models.
We have explored parameter ranges $0.1\le\Omega_M\le1.0$
and $0.3\le\sigma_8\le2.0$.  These constraints are upper limits,
because for fixed \Om it is possible to reduce the number of clusters
by lowering $\sigma_8$.  As expected, high \s8 can be ruled out.
We fit the $2\,\sigma$ contours to give a rough constraint on $\sigma_8$
(for $0.1 \le \Omega_M \le 1.0$):
$\sigma_8 < 1.00\, \Omega_M^{-0.43 \Omega_M-0.22}$ (flat) and
$\sigma_8 < 1.01\, \Omega_M^{-0.40 \Omega_M-0.14}$ (open) at $95\%$ confidence level.
In calculating these results, the uncertainties in detection efficiency
are included (see \S\ref{sec:cobe} for a detailed discussion).

\myputfigure{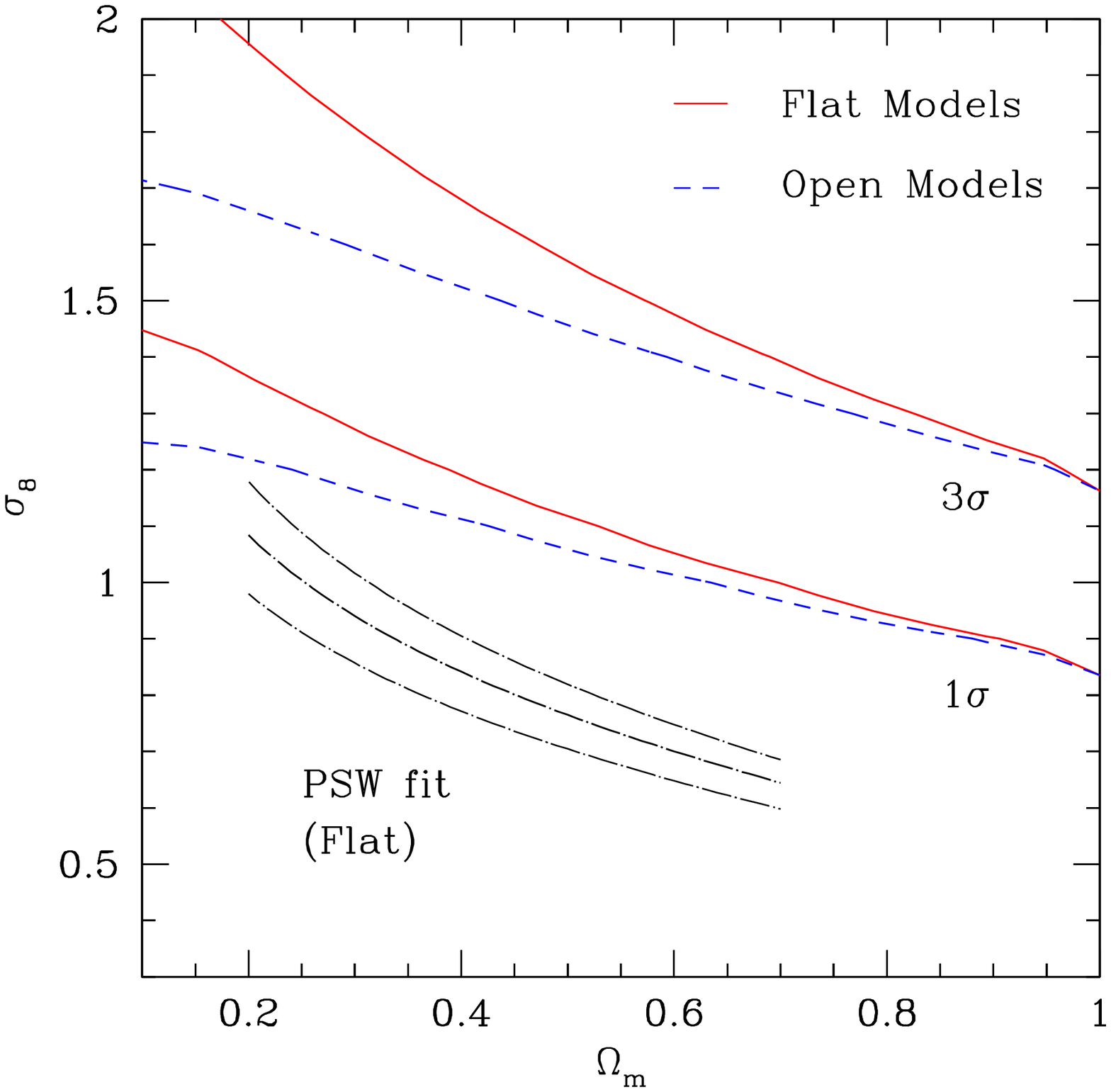}{3.2}{0.5}{-30}{-0}
\figcaption{Constraints on $\Omega_M$ and $\sigma_8$
for both flat (solid line) and open
(dashed line) universe models.
The range of \Om considered is $0.1 \le \Omega_M \le 1.0$, and
the range of \s8 is $0.3 \le \sigma_8 \le 2.0$.
Results from the Pierpaoli et al. (2001) analysis of the temperature
function are shown (dot-dashed line) with $\pm 1\sigma$ range.
\label{fig:bima}}
\bigskip

For comparison, dot--dashed lines in Fig~\ref{fig:bima} 
indicate the preferred values and
1$\sigma$ confidence regions derived from the local temperature function
of 38 massive clusters (\citealt{pierpaoli01}, hereafter PSW).  The PSW
constraints are only for flat cosmologies.  These constraints are clearly
much stronger than those from the BIMA survey.  A larger solid angle
and higher sensitivity SZE survey which delivers SZE cluster detections,
is required to move from upper limits to a preferred range of values, as in
the PSW analysis.  These two surveys -- a large solid angle local survey and
a small solid angle high redshift survey -- are complementary (rather than
redundant); in $\S$\ref{sec:local} we explore joint constraints of the deep
BIMA survey and a similar but simpler approach to the local abundance analysis.

We can understand the shape of the confidence region as follows.
For fixed $\Omega_M$, models of higher \s8 lead to larger $\langle N \rangle$,
which are more inconsistent with the BIMA survey.
The behavior of the probability along lines of fixed $\sigma_8$ is
explained with the help of Fig~\ref{fig:exp}.
In the figure we consider three flat models and one open model
with $\sigma_8=1$:
flat $\Omega_M=0.1$ (solid/red), flat $\Omega_M=0.5$ (dashed/blue),
flat $\Omega_M=1$ (dotted/black) and open $\Omega_M=0.5$ (dot-dashed/green).
From top to bottom the panels contain the cluster redshift distribution
$dN/dz\,d\Omega$, the comoving differential
volume element $dV/dz\,d\Omega = c\,d_A^2 (1+z)^2/H(z)$, the detectable
cluster number density or abundance
$n = \int_{0}^{\infty}\,dM\,dn/dM f(M)$, and the limiting
detectable mass \mlim.  In the case of the limiting mass, we show the ratio
of \mlim in each model to its value in the $\Omega_M=1$ model.
The curves in the top panel are essentially the
products of those in the second and third panels. The behavior of the volume
element reflects that of the angular diameter distance, which changes
only slowly beyond $z\sim1$.  At $z>1$, the BIMA survey is
probing an order of magnitude higher volume in the $\Omega_M=0.1$ model
than in the $\Omega_M=1$ model.

The behavior of the abundance evolution in each model is
primarily a reflection of the way $\sigma(M, z=0)$ is
normalized:  although $\sigma_8$ is set to be unity for all three models, 
the mass contained in $8h^{-1}$~Mpc scale ($\equiv M_{cl}$) differs
for each model because $M_{cl}\propto \Omega_M$. 
The result is that, at low mass range $M\gtrsim M_{lim}$, $\sigma(M,z)$
of high $\Omega_M$ models is always larger.
Interestingly, for masses near the detection limit the reduction in the
amplitude of $\sigma(M)$ for low $\Omega_M$ models at fixed $\sigma_8$
more than compensates for the more rapid evolution of density perturbations
in the high $\Omega_M$ models.  Furthermore, the steepness of the
mass function implies that contributions
to the number density come mostly from the systems with masses near the
detection limit.  Finally, the limiting masses for low density models are 
larger than that of high density models.  Together, these effects lead to 
lower number densities of detectable clusters in the low $\Omega_M$ models.

Finally, consider why the BIMA survey constrains open models more strongly
than flat models. As also shown in Fig~\ref{fig:exp}, the dot-dashed/green
line represents
an $\Omega_M=0.5$ open model. We see there is little difference in the volume
element between two $\Omega_M=0.5$ models, while the difference in abundance at
high redshifts is large. This can be understood from the slower growth of
structure in open models;
in short, open models at fixed $\sigma_8$ produce
more high redshift clusters than flat models
and are thus more inconsistent with the data.

\subsection{BIMA Deep Fields Plus Local Abundance}
\label{sec:local}

In this case, we use both the BIMA deep fields and a measurement
of the local massive cluster abundance.
Calculating the constraints for each \Om requires two steps:  (1) determining
the $\sigma_8$, which reproduces the local abundance of massive clusters above
some fiducial temperature $kT_X$, and (2) then estimating the expected number
of clusters $\langle N \rangle_i$ for each BIMA field $i$.
We determine \s8 by requiring that $n(M>M_{th})$ matches the observed local
cluster abundance, where $M_{th}$ is the mass corresponding to the X--ray
temperature cutoff of the sample. 
We adopt the value given by \citet{viana99}, which is for
clusters of \xray temperature greater than $6.2\kev$ at $z=0.05$
(see also \citealt{henry97}).
The values of \s8 we obtain do not differ much from those of
\citet{pierpaoli01}, within the range of $0.2 \le \Omega_M \le 0.7$.
We then calculate $\langle N \rangle_i$ in each BIMA field and the
corresponding probability for that cosmological model.

However, combining these two datasets provides very weak, if any, constraints
on the density parameter; essentially all models considered are consistent with
the data. Models with lower matter density are relatively more incompatible,
but even at the model with lowest density ($\Omega_M = \Omega_B$), the
exclusion probability is still weaker than $1\sigma$.

\myputfigure{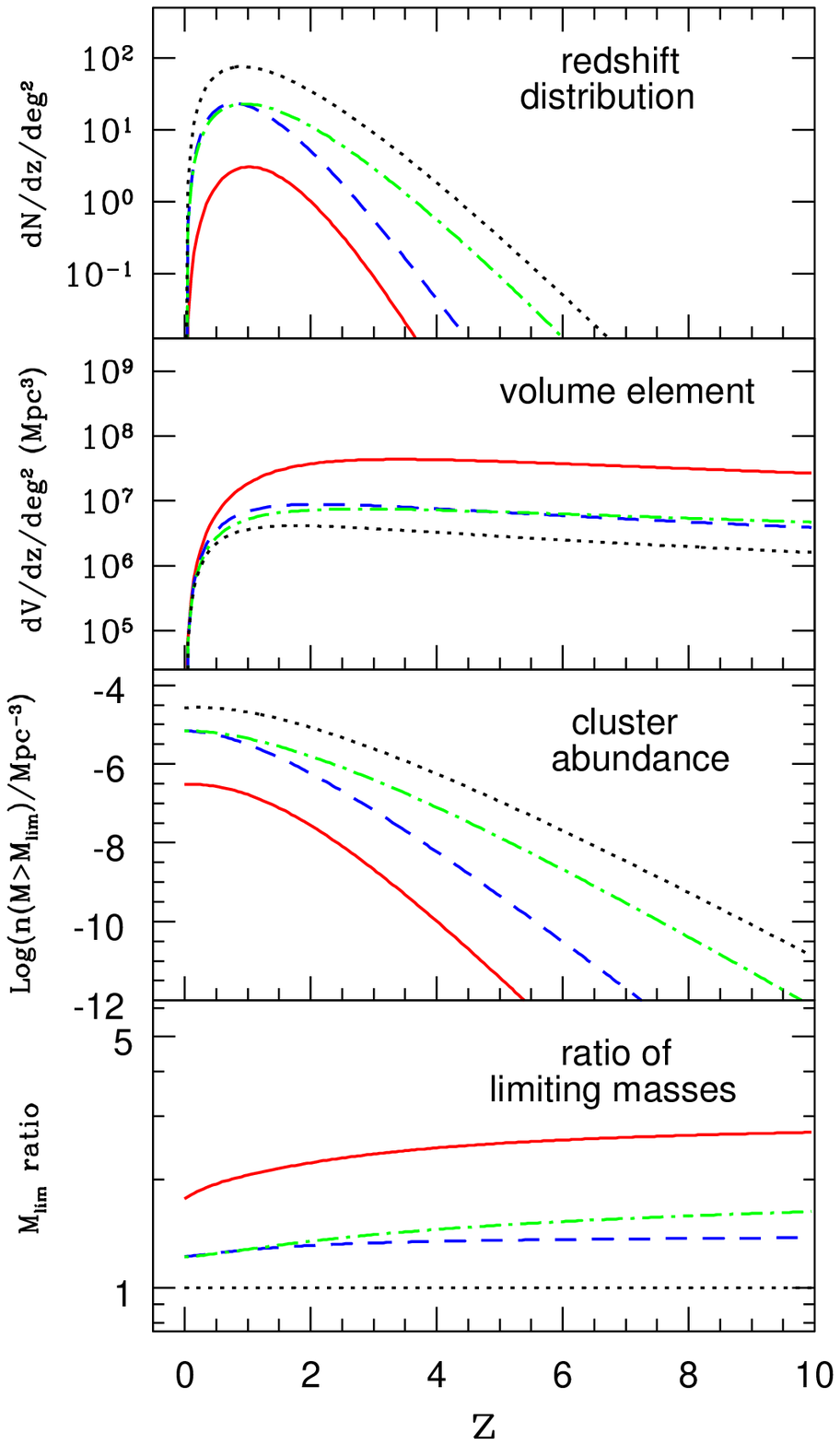}{3.2}{0.52}{-20}{-0}
\figcaption{The behavior of the confidence regions in
Fig~\ref{fig:bima} results from several competing effects.
From top to bottom, we show the cluster redshift distribution, the
volume element, the cluster abundance, and the limiting
mass for three flat models at fixed $\sigma_8=1$ and $h=0.65$: 
$\Omega_M=0.1$ (solid), $\Omega_M=0.5$ (dashed), and $\Omega_M=1$ (dotted).
Also shown is an $\Omega_M=0.5$ open model (dot--dashed).  The limiting mass
panel contains the ratio of $M_{lim}$ to that in the $\Omega_M=1$ model.
\label{fig:exp}}
\bigskip

This method produces a {\it lower} limit on $\Omega_M$
because we are requiring the local cluster abundance to be the same in all
cosmological models;  in this scenario, those models with slower
growth of density perturbations (i.e. low $\Omega_M$ models) produce
more high redshift clusters and are therefore more inconsistent with
our SZE survey.

\subsection{BIMA Deep Fields Plus COBE}
\label{sec:cobe}

By combining the BIMA deep fields with a COBE normalization for
the power spectrum of density fluctuations, it is possible to
place an upper limit on \Ome.  The COBE measurements on
large angular scale anisotropy can be used to constrain
the amplitude of the power spectrum on the scale of the present 
epoch horizon ($3000 \sim 4000\, h^{-1}$ Mpc). 
Taking the amplitude of modes on this scale to be 
$\delta_H$, we can write the power spectrum in the form
$P(k,z) = 2\pi^2\,\delta_H^2\,(c/H_0)^{3+n}\,k^n\,T^2(k)\,D^2(z)/D^2(0)$
\citep[e.g.][]{bunn97,eisenstein98}.
We adopt the fitting function for parametrization of $\delta_H$ 
provided by \citet{bunn97}, assuming no contributions from primordial
gravitational waves. For the index of primordial power spectrum, we
adopt the BOOMERANG measurement $n=0.96^{+0.10}_{-0.09}$ \citep{netterfield01}.
Although we adopt different values of $\Omega_B$ and $H_0$ than those used
by \citet{bunn97}, we do not rescale $\delta_H(\Omega_M, n)$ to account
for these differences; the change in amplitude of $\delta_H$ from this
rescaling would be small compared to the statistical uncertainty on $\delta_H$ 
itself.  Normalizing the power spectrum
in this way leads to \s8 that increases with \Ome.

Because \s8 increases with \Ome, we expect the COBE normalized SZE
survey constraints to provide an upper limit on \Ome.
From Fig~\ref{fig:cobe}, we find that the 95\% confidence upper
limit on $\Omega_M$ in flat (open) models is $\Omega_M<0.63$ ($\Omega_M<0.73$).
These upper limits include uncertainties in the SZE flux--virial mass relation,
the Hubble parameter, the COBE normalization and the power spectrum index, as well
as the effect of the sample variance.  Below we provide further detail.

\myputfigure{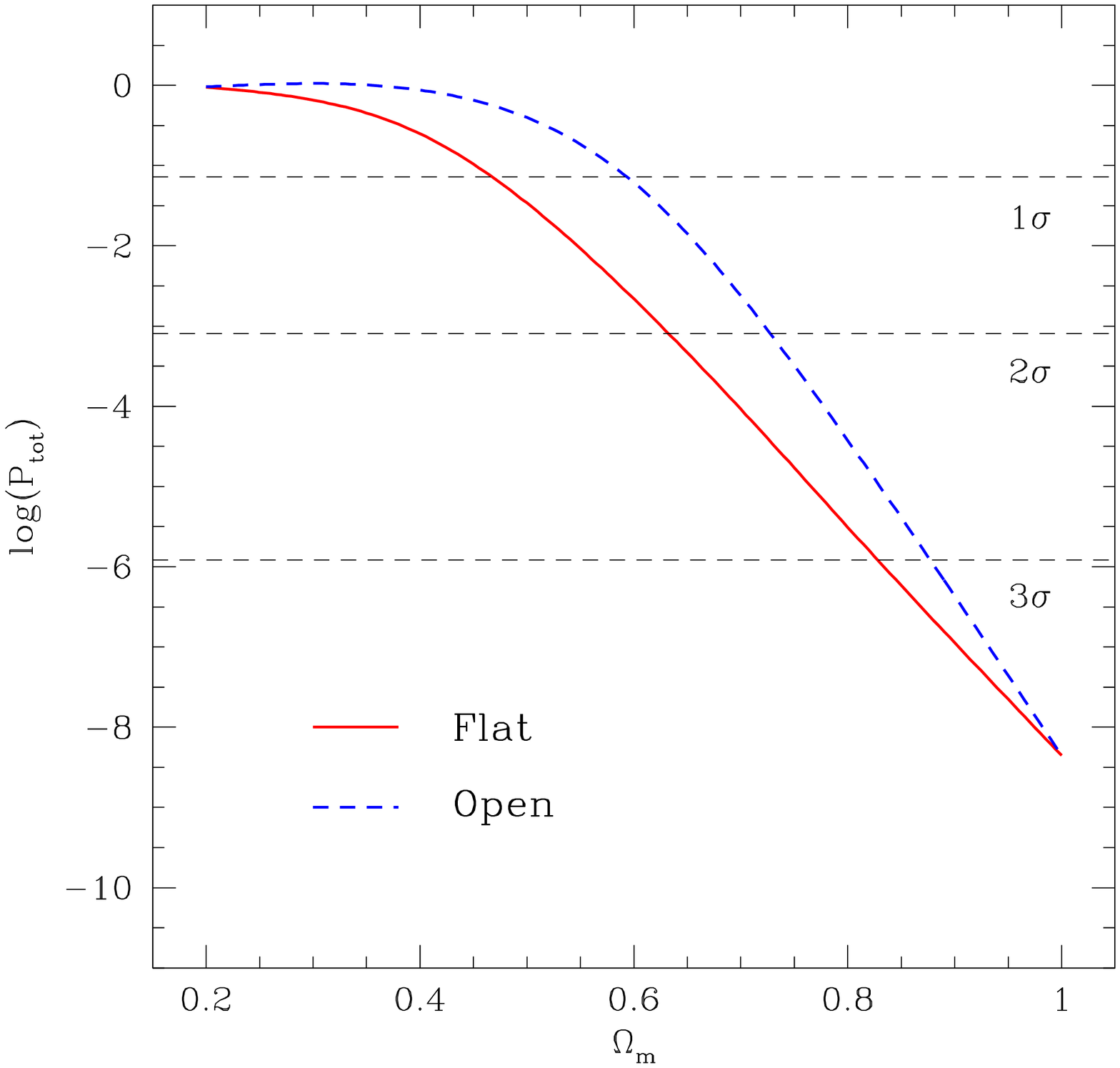}{3.2}{0.5}{-80}{-30}
\figcaption{Exclusion probability as a function
of $\Omega_M$ for both flat and open models,
when the power spectrum is normalized to COBE.
Uncertainties in virial mass, $H_0$, $\delta_H$, and $n$ as well as the sample variance
are included (see \S\ref{sec:cobe}).
In flat (open) models, $\Omega_M<0.63$ (0.73) at 95\% confidence.
\label{fig:cobe}}

\subsubsection{Detection Efficiency}

The key to interpreting cluster surveys is connecting the observables 
(cluster SZE visibility in this case) to the halo virial mass.  This
generally involves a mass--observable scaling relation, which we term the
visibility--mass relation -- and a model for the redshift evolution
of this relation.  As has already been discussed, we use hydrodynamical
simulations to define this visibility--mass relation, and we parametrize
the redshift evolution of this visibility--mass relation by making mock
observations of clusters output at a range of redshifts.  The
visibility--mass relation for an interferometric SZE survey is similar to
the better known mass--temperature relation in an X-ray cluster survey.
In the X-ray survey the cluster temperature is used as the mass predictor.

We model these uncertainties by introducing a 40\% systematic uncertainty
in the normalization of the $\Delta\chi^2$--mass 
relation (see Fig~\ref{fig:deltachi}), corresponding
to a 20\% uncertainty in the visibility--mass relation.  
From Eqn~\ref{eq:flux} it is clear that this corresponds to a systematic
uncertainty in the limiting cluster mass (remember $f(M_{lim})\equiv0.5$)
of $\sim 16\%$ (see the discussion in \S\ref{sec:hscaling}). 
We model this systematic uncertainty as a Gaussian distribution, and 
marginalize over this systematic by integrating the probability
obtained from Eqn~\ref{eq:comb}
over the Gaussian distribution in $M_{lim}$.

\subsubsection{Hubble Parameter $H_0$ and 
Power Spectrum Parameters $\delta_H$ and $n$}

Our constraints on $\Omega_M$ are sensitive to the Hubble parameter through its effect on the volume element, the amplitude of $\sigma(M)$, the critical
density and the mass limit. 
Rather than choosing a particular value of $H_0$, 
we assume a Gaussian distribution in $H_0$ centered on a fiducial value of
65~km/s/Mpc with a 1$\sigma$ uncertainty of $10\%$.
We also account for uncertainties in the COBE normalization
amplitude $\delta_H$ and the primordial power spectrum index $n$.
We take both the distributions of $\delta_H$
and $n$ to be Gaussian, with dispersions of $7\%$ \citep{bunn97}
and $10\%$ \citep{netterfield01}, respectively.  The final probability of consistency with
the SZE survey is calculated by integrating over these Gaussian distributions.

\subsubsection{Sample Variance}
\label{sec:variance}

To assess the importance of sample variance on our $\Omega_M$ constraints,
we follow the approach of \citet{hu02}.
Because our survey fields are small, we approximate
each field as a series of circular pillboxes, each of radius $3'_.5$ and thickness $\Delta z=0.16$, arranged in increasing redshift. We calculate the covariance matrix $\sigma^2_{ij}$  between redshift bins $i$ and $j$
\begin{eqnarray}
\sigma^2_{ij} &= & n_i \overline{b}_i \ n_j \overline{b}_j
D_i D_j \int \frac{d^3 k}{(2\pi)^3}W_iW_j^* P(k),
\end{eqnarray}
where $n$ is the number of detected clusters in the bin, $\overline{b}$ is the mass function weighted mean bias, $D$ is the growth function and $W$ is the Fourier transform of the pillbox window function at each redshift.
We estimate the sample variance $\sigma_s^2$ by summing 
the matrix elements.  We have only included the 
terms with $|i-j|\le 1$, because other elements are negligible, as noted 
by \citet{hu02}. The exclusion probability for a given cosmology is then
\begin{equation}
P(\Omega_M) \propto \int_0^\infty\!\!\! d\mu\,e^{-{(\mu-\langle N \rangle)^2 \over 2 \sigma_s^2}}
	\,e^{-\mu}
\end{equation}
up to a suitable normalization factor. It turns out, the inclusion of the
sample variance only weaken our constraints to a small extent, corresponding roughly
to a $1\%$ upward shift in the constraint curve in Fig~\ref{fig:cobe}.

\subsubsection{Field Selection and Non--detection Significance}
\label{sec:select}

These fields were not chosen randomly, and the selection criteria could potentially bias 
the expected number of detected clusters.
Fields p1643 and VLA1312 are centered in directions claimed to
have cluster SZE decrements \citep{jones97,richards97} that have not been
 confirmed by higher sensitivity observation \citep{holzapfel00b}.  The field pss0030 is in the direction of a 
radio--quiet quasar. The other 4 fields were chosen to be evenly distributed
in right ascension, at convenient declinations, in regions of low dust--emission 
and free of bright point sources at $1.4$ GHz, in optical and in X--ray. 
As pointed out by \citet{dawson01}, radio
point sources may trace large scale structure, preferentially populating
clusters.  If so, as a whole the fields may be biased to contain fewer than the expected number of clusters.

Clearly, our constraints on $\Omega_M$ are valid only if these fields provide an unbiased sample of the universe.  One could also ask whether it is surprising that no clusters were detected in the deepest and largest SZE survey to date.  If the fields are unbiased, then the answer is no, if we live in a low matter density universe.  If the fields are biased against finding clusters, then this would make it even less surprising that no clusters were detected.

Our results also depend to some degree on our interpretation of the \citet{holzapfel00a} search for cluster SZE decrements.  Specifically, we present results assuming that no clusters were detected at or above the 95\% confidence level, whereas the authors provide no quantitative discussion.  Clearly, the authors were very interested in finding galaxy clusters, and no firm detections were made \citep{holzapfel00a,holzapfel00b}.  A reanalysis of the data searching for clusters in much the same way that we have searched for clusters in our mock observations would be ideal, and such an analysis is ongoing (Carlstrom, private communication).  Here we simply report the constraints on $\Omega_M$ if cluster detections in the BIMA deep fields can only be ruled out at the 3$\sigma$ and 4$\sigma$ level.  In these cases, clusters would have to be more massive to produce these stronger detections, and therefore one would generally expect fewer detections for a given density parameter.  Thus, the 95\% confidence upper limit would weaken to $\Omega_M<0.70$ and $\Omega_M<0.77$, respectively.  In general, our results are relatively insensitive to the exact characterization of the non--detection significance.

\section{Summary and Discussion}
\label{sec:summary}

We describe a systematic analysis of an interferometric SZE survey carried 
out at BIMA with the Carlstrom--Joy 30~GHz receivers. The data 
were taken from observations of the CMB anisotropy on arcminute scale 
($l \sim 5500$), in which no anisotropy and no galaxy clusters were detected 
\citep{holzapfel00a,holzapfel00b}.   We use these data to study
the allowed range of cosmological parameters $\sigma_8$ and $\Omega_M$.
In our analysis, we consider models that are flat with a cosmological
constant ($\Omega_M+\Omega_\Lambda=1$) and models that are open with
no cosmological constant. 

Our analysis hinges on our ability to calibrate the survey sensitivity or
galaxy cluster selection function.  To this end, we use mock interferometric
observations of hydrodynamical cluster simulations to determine the selection
function for each of the seven fields and as a function of redshift and
position within that field.  Relying on hydrodynamical simulations to calibrate
the survey sensitivity introduces significant uncertainties, which can
be reduced once empirical calibration using large surveys that 
include detailed followup observations are available.  For this analysis we 
estimate the scale of the uncertainties that come with using hydrodynamical cluster simulations and include those uncertainties in determining our 
constraints on the cosmological parameter \Ome.

With BIMA survey data alone, 
we obtain a 95\% confidence upper limit on $\sigma_8(\Omega_M)$: 
$\sigma_8 < 1.00\, \Omega_M^{-0.43 \Omega_M-0.22}$ for flat, and
$\sigma_8 < 1.01\, \Omega_M^{-0.40 \Omega_M-0.14}$ for open models.
Combining the BIMA survey with external constraints on the power spectrum
allows us to constrain \Om alone.  In combination with the local
abundance of high temperature X--ray clusters, the BIMA data lead only to a trivially
weak lower limit: $\Omega_M>\Omega_B$.
In combination with the COBE normalization of the power spectrum, 
the data provide an upper limit on $\Omega_M$: 
$\Omega_M<0.63$ for flat models, and $\Omega_M<0.73$ for open models at
95\% confidence.   Included in the constraints on \Ome are 40\% uncertainties in the cluster visibility--mass relation as well as published uncertainties on the Hubble parameter and power spectrum normalization $\delta_H$ and spectral index $n$.  We also include the effects of sample variance.

These constraints are valid only if these BIMA fields represent an unbiased sample of the universe.   The field selection in this survey is somewhat suspect, with four of the seven fields specifically chosen to exclude bright radio, X--ray and optical point sources.  A firm interpretation of our results is that --
if the fields are an unbiased clusters survey -- we have shown that within the context of the cosmological models favored today, we expect no clusters to have been detected.  If the survey was biased against finding clusters, then our constraints on \Om weaken, and it is even less surprising that no clusters were detected.  Of course, if the survey were biased toward finding clusters, then finding none would imply tighter constraints on \Ome.  We eagerly await planned SZE surveys of larger and perhaps better chosen fractions of the universe.

In previous work, \citet{majumdar00} studied constraints on $\Omega_M$
by comparing simulated SZE maps with the upper limit on the 
arcminute--scale anisotropy in CMB obtained by \citet{subrahmanyan00} using
the Australia Telescope Compact Array (ATCA) at $8.7$ GHz. Their 95\% 
confidence upper limit is $\Omega_M<0.8$ for open models. 
This result is clearly
consistent with ours, but there are important differences in the analysis and
the data.  First, the ATCA survey has lower sensitivity than the BIMA survey,
yielding a 95\% confidence limit of $Q_{flat} < 25\,\mu$K compared to
the BIMA survey constraints of $Q_{flat} < 14.1\,\mu$K.  Second, their focus
was on reproducing the statistical properties of the noise in the sky maps, 
whereas we have on the non--detection of clusters. 
Third, our modeling technique is more sophisticated than theirs through its use of hydrodynamical cluster simulations that incorporate the effects of cluster morphology and internal structure and the evolution of these properties with redshift.
And fourth, we model the redshift evolution of the cluster abundance using
the results from the latest N--body studies of structure formation 
\citep{jenkins01}, whereas they modeled cluster abundance using the original
Press-Schechter formalism.  We believe that the techniques we outline in this manuscript
provide a viable approach for analyzing the forthcoming interferometric SZE
surveys that will use much higher sensitivity instruments to survey larger
portions of the sky.

In short, the principle improvement of our analysis over previously
existing analyses and estimates of survey yields include (1) the direct
accounting for the uncertainty in the cluster detectability (specifically,
we model the scatter in the detection significance--virial mass relation and we include
the uncertainty in the normalization of the SZE visibility--virial mass 
relation), (2) the modeling of the survey sensitivity as a function
of position on the sky, (3) the inclusion of uncertainties on other cosmological parameters of interest, and (4) an accounting for the effects of sample variance on our \Om constraint.  

In addition, we have examined the $h$ dependence of 
the survey limiting mass (see Appendix), showing that for this interferometric survey, which
includes sensitivity to cluster shape {\it and} flux, the $h$ dependence departs
from the flux limited survey expectation $M_{lim}\propto h^{-8/5}$ to a steeper
dependence of $M_{lim}\propto h^{-11/6}$.  Similar departures would be expected for other interferometric surveys.

Large solid angle, high redshift SZE cluster surveys can in
principle provide a wealth of information on structure evolution and cosmology.
Because of its redshift--independent nature, the SZE is a powerful tool 
in studies of the high redshift universe. An interferometric
SZE survey has advantages such as superb angular resolution and the ability to
separate point source contamination from the extended clusters 
\citep{holder00,kneissl01}. Future SZE surveys will have much higher 
sensitivity than the BIMA survey analyzed here, and 
will be carried out with the Sunyaev-Zel'dovich effect Array \citep[SZA,][] 
{holder01b}, the Arcminute Microkelvin Imager \citep[AMI,][] {kneissl01}, and 
the Arcminute Mm-wave interferometer for Background Anisotropy 
\citep[AMiBA,][]{lo00}.   These surveys are expected to detect hundreds of 
clusters, and these large, high redshift cluster samples will be 
invaluable to cosmological studies, enabling tests of the mix of dark energy and dark matter in our universe and measurements of the nature of this dark energy\citep[e.g.][]{barbosa96,haiman01,holder01b,fan01,mohr01,weller01,hu02}.

\acknowledgements
We acknowledge many illuminating conversations with Subha Majumdar.  
We thank Bill Holzapfel and John Carlstrom for providing details
about the BIMA deep field observations, and we thank Erik Reese, Marshall Joy and 
Laura Grego for taking part in the observations that began this project.  
We acknowledge Gus Evrard for providing the hydrodynamical cluster simulations
used to determine the survey sensitivity.  YTL acknowledges financial support from 
the University of Illinois Research Board.  JJM acknowledges financial support from 
the NASA Long Term Space Astrophysics grant NAG 5-11415.

\bibliographystyle{../Bib/Astronat/apj}
\bibliography{../Bib/cosmology}

\appendix
\label{sec:hscaling}

Here we examine the $H_0$ dependence of survey sensitivity, showing how in an interferometric survey it departs significantly from the expected scaling for a flux limited survey.  For an approximately virialized and self--similar cluster population, 
the density weighted electron temperature scales with the cluster mass: 
$\left<T_e\right> \propto (M h)^{2/3}$ \citep[e.g.][]{bryan98,mohr00a}. 
Eqn~\ref{eq:flux} then implies that the galaxy cluster SZE flux scales as 
$S_\lambda \propto M_{200}^{5/3} h^{8/3}$.
Therefore, the limiting mass in an SZE flux limited survey would scale as 
$M_{lim} \propto h^{-8/5}$.

In interferometric surveys, it is the cluster apparent size and morphology
along with the total cluster flux that play important roles in determining
the detectability of a cluster.  The limiting mass is that mass which produces a particular detection significance $\Delta\chi^2$, where the detection significance 
scales as the cluster visibility squared.   From Eqn~\ref{eq:VandS} it is clear 
that cluster visibility is proportional to the cluster flux, so if flux were 
the only important factor we would expect 
$\Delta \chi^2 \propto M_{200}^{\alpha}$ with $\alpha=10/3$. However, the best 
fit $\Delta \chi^2$ -- $M_{200}$ relations exhibit significantly shallower 
slopes: $a \sim 2.3 - 2.4 < 10/3$ (at $z=0.5$; see Fig~\ref{fig:deltachi}). 
This underscores the sensitivity of the visibility to cluster angular size and 
shape.  At a particular redshift, increasing cluster mass means increasing total cluster flux, but this effect is partially offset by the larger cluster apparent size, which reduces the signal at any particular baseline.  

There is also considerable scatter in the $\Delta \chi$--$M_{200}$ relation.  
We have inspected the simulated clusters that fall well below the 
best fit $\Delta \chi^2$ -- $M_{200}$ relation; these clusters
generally appear more extended and irregular (in some cases bimodal). 
Many of these morphologically complex clusters are obviously in the midst
of a major merger, and their complex shape causes diminution of the cluster
visibility over the angular scales where the BIMA survey is most sensitive.

To obtain the $h$ dependence of the interferometric survey sensitivity, we need only know the
slope of the $\Delta \chi^2$--$M_{200}$ relation and the $h$ dependence
of $\Delta \chi^2$ and $M_{200}$.  At fixed ICM temperature $\left<T_e\right>$,
$M_{200}\propto h^{-1}$ and $\Delta \chi^2\propto h^2$.  Including the
$h$ dependence in the $\Delta \chi^2$--$M_{200}$ relation
leads to $\left(\Delta \chi^2h^{-2}\right)\propto \left(M_{200}h\right)^\alpha$.  Keeping
the required detection significance $\Delta \chi^2$ fixed provides $h$
scaling: $M_{lim}\propto h^{-(2+\alpha)/\alpha}$.  
For $\alpha=10/3$ which is derived from assuming only flux is important in 
determining the cluster detection significance,
we arrive at $M_{lim}\propto h^{-8/5}$, consistent with expectations for a flux
limited survey. Taking the typical measured slope $\alpha=2.4$, we calculate 
the true $h$ sensitivity of the limiting mass to be $M_{lim}\propto h^{-11/6}$.
The shallower the $\Delta \chi^2$--$M_{200}$ relation, the larger shift in
$M_{lim}$ required to offset a change in $h$.

\end{document}